%% file: zeynep_archive_2016.tex
\documentclass[final] {aipproc}
\usepackage{sidecap,graphicx,boxedminipage,epsfig,wrapfig,floatflt,shadow}

\layoutstyle{8x11double}

\begin{document}

\title{Improving Student Understanding of Coulomb's Law and Gauss's Law}
\classification{01.40Fk,01.40.gb,01.40G-,1.30.Rr}
\keywords      {physics education research}

\author{Zeynep Isvan and Chandralekha Singh}{
  address={Department of Physics and Astronomy, University of Pittsburgh, Pittsburgh, PA, 15260, USA}}

\begin{abstract}
We discuss the development and evaluation of five research-based tutorials on
Coulomb's law, superposition, symmetry and Gauss's Law to help students in
the calculus-based introductory physics courses learn these concepts.
We discuss the performance of students on the pre-/post-tests given before
and after the tutorials in three calculus-based introductory physics courses.
We also discuss the performance 
of students who used the tutorials and those who did not use it
on a multiple-choice test which employs concepts covered in the tutorials.

\end{abstract}

\maketitle

\section{Introduction}

Electrostatics is an important topic in most calculus-based introductory physics courses.
Although Coulomb's law, superposition principle, and Gauss's law are taught in
most of these courses, investigations have shown that these concepts are challenging
for students~\cite{maloney,rainson,singh}. Despite the fact that students may have learned the
superposition principle in the context of forces in introductory mechanics, this learning
does not automatically transfer to the abstract context of electrostatics and students get
distracted by the very different surface features of the electrostatics problems.
Effective application of Gauss's law implicitly requires
understanding the principle of superposition for electric fields and the symmetry
that ensues from a given charge distribution.
Helping students learn these concepts will not only help them build a more coherent
knowledge structure, it can also improve their reasoning and meta-cognitive skills.
Here, we discuss the development and evaluation of research-based tutorials and the corresponding 
pre-/post-tests to help students develop a functional understanding of these concepts.

\vspace{-.05in}
\section{Tutorial Development and Administration}

\input All_Q_by_Q_Rounded1_new

Before the development of the tutorials, we conducted investigation of student difficulties
with these concepts~\cite{singh} by administering free-response and multiple-choice questions and by
interviewing individual students. 
We found that many students have difficulty distinguishing between the electric charge, field and force.
Students also have difficulty with the principle of superposition and in recognizing whether sufficient
symmetry exists for a particular charge distribution to calculate the electric field using Gauss's law.
Choosing appropriate Gaussian surfaces to calculate the electric field using Gauss's law
when sufficient symmetry exists is also challenging for students.
Distinguishing between electric field and flux was often difficult.

\input All_Rounded_new

We then developed the preliminary version of five tutorials and the corresponding pre-/post-tests
based upon the findings of the difficulties elicited in previous research and a theoretical task analysis of the underlying concepts.
Theoretical task analysis involves making a fine-grained flow chart of the concepts involved in solving specific class of problems.
This type of analysis can help identify stumbling blocks where students may have difficulty.
The first two tutorials were developed to help students learn about
Coulomb's law, superposition principle and symmetry in the context of discrete and continuous charge distributions (conceptually),
the third tutorial focused on distinguishing between electric flux and field, and the fourth and fifth
tutorials dealt with symmetry and Gauss's law and on revisiting superposition principle after Gauss's law.
Although some tutorials on related topics have been developed by the University of Washington group, those tutorials
are complementary to the ones we have developed focusing on symmetry ideas.
We administered each pre-test, tutorial and post-test to 5 students individually
who were asked to talk aloud while working on them. After each administration, we modified the
tutorials based upon the feedback obtained from student interviews. These individual administrations 
helped fine-tune the tutorials and improve their organization and flow. Then, the tutorials
were administered to four different calculus-based introductory physics classes with four lecture hours and one recitation
hour per week. Students
worked on each tutorial in groups of two or three either during the lecture section of the class or in the
recitation depending upon what was most convenient for an instructor. Table 1 shows the pre-/post-test data
on each question from three of the classes in which the tutorials were administered. The details of each question will be discussed
elsewhere. In the fourth class, the post-tests were
returned without photocopying them and we only have complete data on student performance on the cumulative test
administered after all tutorials. As shown in Table 1,
for some tutorials, additional questions were included in the pre-test and/or post-test after the previous
administration and analysis of data. The pre-/post-tests were not identical but focused on the same topics covered in a tutorial.

\input Gauss_Law_Avg_short

All pre-tests and tutorials were administered after traditional instruction in relevant concepts. Instructors
often preferred to alternate between lectures and tutorials during the class and give an additional
tutorial during the recitation. 
This way all of the five tutorials from Coulomb's law to Gauss's
law were administered within two weeks. For the tutorials administered in lecture section of the class, pre-tests were given to students right before
they worked on the tutorials in groups. Since not all students completed a tutorial during the class,
they were asked to complete them as part of their homework assignment. At the beginning of the next
class, students were given an opportunity to ask for clarification on any issue related to the part of the tutorial
they completed at home and then they were administered the corresponding post-test before the lecture began. Each pre-/post-test
counted as a quiz and students were given a full quiz grade for taking each of the pre-test regardless of students' actual performance. 
The pre-tests were not returned but the post-tests were returned after grading. 
When a tutorial was administered in the recitation (the second and fifth tutorials which were shorter), 
the teaching assistant (TA) was given specific 
instruction on how to conduct the group work effectively during the tutorial. Moreover, since the TA had to 
give the post-test corresponding to the tutorial during the same recitation class in which the students worked
on the tutorials (unlike the lecture administration in which
the post-tests were in the following class), the pre-tests were skipped for some
of these tutorials due to a lack of time. 
Sometimes, the instructors gave the pre-tests 
in the lecture section of the class for a tutorial that was administered in the recitation.

In all of the classes in which the tutorials were used, 2-2.5 weeks were
sufficient to cover all topics from Coulomb's law to Gauss's law.
This time line is not significantly different from what
the instructors in other courses allocated to this material. The main difference between the tutorial
and the non-tutorial courses is that fewer solved examples were presented in the tutorial classes (students
worked on many problems themselves in the tutorials). We note that since many of the tutorials were administered
during the lecture section of the class, 
sometimes two instructors (e.g., the instructor
and the TA) were present during these ``large" tutorial sessions to ensure smooth facilitation.
In such cases, students working in groups of three were asked to raise their hands for questions and clarifications.
Once the instructor knew that a group of students was making good progress, that group was invited to help other
groups in the vicinity which had similar questions. Thus, students not only worked in small groups discussing
issues with each other, some of them also got an opportunity to help those in the other groups.

\vspace*{-.1in}
\section{Discussion}
\vspace*{-.05in}

Out of the five tutorials that students worked on, the first two focused on Coulomb's law, superposition
and symmetry. The first tutorial started with the electric field due to a single point charge
in the surrounding region and then extended this discussion to two or more point charges. The second tutorial further
continued the conceptual discussion that started in the first tutorial (which was mainly about discrete charges) to continuous charge
distributions. The tutorials guided students to understand the vector nature of the electric field,
learn the superposition principle and recognize
the symmetry of the charge distribution. Students worked on examples in which the symmetry of the charge distribution
(and hence the electric field) was the same but the charges were embedded on objects of different shapes
(e.g., four equidistant charges on a plastic ring vs. a plastic square). Common misconceptions were explicitly
elicited often by having two students discuss an issue in a particular context. 
Students were asked to identify the student with whom they agreed.

The third tutorial was designed to help students learn to distinguish between the electric field and flux. The tutorial
tried to help students learn that the electric field is a vector while the electric flux is a scalar. Also, electric
field is defined at various {\it points} in space surrounding a charge distribution while the electric flux is always
through an {\it area}. Students learn about Gauss's law and how to relate the flux through a closed surface to the net
charge enclosed. 
Rather than emphasizing the symmetry considerations, this tutorial focused on helping students use Gauss's law to find 
the net flux through a closed surface given the net charge enclosed and vice versa. 

The fourth tutorial was designed to help students learn to exploit Gauss's law to calculate the electric field at a point
due to a given charge distribution if a high symmetry exists. Students were helped to draw upon the superposition and symmetry ideas they
learned in the first two tutorials to evaluate whether sufficient symmetry exists to exploit Gauss's law to calculate the
electric field. Then, students learn to choose the appropriate Gaussian surfaces that would aid in using Gauss's law
to find the electric field. Finally, they use Gauss's law to calculate the electric field in these cases. The last
tutorial revisits the superposition principle after students have learned to exploit Gauss's law to calculate the electric field.
For example, students learn to find the electric field at a point due to two non-concentric uniform spheres of charge or due to
a point charge and an infinitely long uniform cylinder of charge. 

The pre-tests and post-tests were graded by two individuals and the inter-rater reliability is good.
The average pre-/post-test scores on matched pairs for a particular class graded by them differed at most by a few percent.
Table 1 shows the student performance (on each question and also overall) on the pre-test and post-test in each of the five tutorials 
(I-V) in percentage. The classes utilizing each tutorial may differ either because additional 
pre-/post-test questions were added or the pre-tests for tutorial II and V were not administered to some of the classes. 
The differences in the performance of different classes may also be due to the differences in student 
samples, instructor/TA differences or the manner in which the tutorials were administered. Table 2 shows the performance of 
students on the pre-/post-tests for each tutorial 
partitioned into three separate groups based upon the pre-test performance (see the Range column). 
As can be seen from Table 2, tutorials generally
helped all students including those who performed poorly on the pre-test.
Table 3 shows the average percentage scores from a cumulative test which includes concepts from all of the tutorials~\cite{singh} 
administered to different student populations. Although the performance of students working on the tutorials is
not as impressive on the cumulative test as on the pre-/post-tests administered with the tutorials, Table 3 shows that students who 
worked through the tutorials significantly outperformed both Honors students and those in upper-level undergraduate courses,
but not physics graduate students.

\vspace*{-.10in}
\section{Conclusion}
\vspace*{-.06in}

We developed and evaluated tutorials to help calculus-based introductory students learn Coulomb's law, 
superposition, symmetry and Gauss's law. Pre-/post-tests for each tutorial and a test that 
includes content on all of the tutorials show that the tutorials can be effective in improving student understanding
of these concepts. 
Moreover, these tutorials appear to be helpful for students who obtained low scores ($0-33\%$) on the 
pretest after traditional instruction.

\vspace*{-.07in}
\begin{theacknowledgments}
We are grateful to the NSF for award DUE-0442087.
\end{theacknowledgments}
\vspace*{-.07in}

\bibliographystyle{aipproc}


\end{document}

%% file: All_Q_by_Q_Rounded1_new.tex



\hspace*{-1.5in}
\begin{table}[h]
\centering
\begin{tabular}[t]{|c|c|c|c|c|c|c|c|c|c|c|c|c|c|c|c|c|c|}
\hline

Tutorial&Class &n&\multicolumn{7}{|c|}{PRETEST}& \multicolumn{7}{|c|}{POSTTEST}\\[0.5 ex]
\cline{4-10}\cline{11-17}
 & & & 1&	2&3&4&	5&	6&Pre-Total& 1&	2&	3&	4&	5&	6&	Post-Total \\[0.5 ex]	
\hline
 &1&82&	        64&	57&	46&	45&	---&	---&	53&92&86&93&---&	---&	---&	90 \\[0.5 ex]
I &2&60&	52&	58&	38&	47&	---&	---&	48&96&96&95&---&	---&	---&	96 \\[0.5 ex]
 &3&52&		44&	29&	45&	---&	---&	---&    39&85&77&88&---&	---&	---&	83 \\[0.5 ex]
\hline
 &1&84&	        ---&	---&	---&	---&	---&	---&	---&68&84&68&72&90&---&	76 \\[0.5 ex]
II &2&63&	56&	6&	41&	---&	---&	---&	35&90&96&87&87&98&	---&	92 \\[0.5 ex]
 &3&63&	        ---&	---&	---&	---&	---&	---&	---&75&84&77&77&92&---&	81 \\[0.5 ex]
\hline
 &1&78&	        42&	---&	---&	---&	---&	---&    42&77&85&72&92&81&---&81	 \\[0.5 ex]
III &2&55&	44&	---&	---&	---&	---&	---&	44&74&88&82&95&---&---&85 \\[0.5 ex]
 &3&49&	        40&	---&	---&	---&	---&	---&	40&81&78&84&96&87&---&	85 \\[0.5 ex]
\hline
 &1&65&	        28&	22&	58&	58&	41&	17 &	40&83&91&95&91&91&93&	91 \\[0.5 ex]
IV &2&62&	39&	19&	51&	52&	42&	6  &	36&85&84&93&96&95&84&	89 \\[0.5 ex]
 &3&49&	        45&	6&	38&	28&	---&	---&	30&87&90&88&81&---&---&	87 \\[0.5 ex]
\hline
 &1&85&	        ---&	---&	---&	---&	---&	---&	---&71&61&75&70&96&64&	71 \\[0.5 ex]
V &2&57&	21&	26&	35&	---&	---&	---&	27&82&76&84&---&---&	---&	80 \\[0.5 ex]
 &3&64&	        ---&	---&	---&	---&	---&	---&	---&92&81&89&89&95&90&	88 \\[0.5 ex]
\hline
\end{tabular}
\vspace{0.1in}
\caption{Average percentage scores obtained on individual questions on the pre-/post-tests (matched unless only the post-test
was given)
for each of the five 
tutorials (I-V). The pre-tests were administered after traditional instruction but before the tutorial.
As shown in the table, additional questions were included either in the pre-test or post-test and the pre-tests for tutorials
II and V were not administered 
in some of the classes. The symbol $n$ refers to the matched number of students in a given class for a given pre-/post-tests
and $Total$ refers to the total average percentage score including all questions on a 
pre-test or post-test administered to a given class for a particular tutorial. 
For tutorial II, the relative weights for the three pre-test questions for class 2 were $30\%$, $30\%$ and $40\%$ 
respectively. 
For tutorial IV, the relative weights for the pre-test and post-test questions for classes 1 and 2 were
$10\%$, $10\%$, $20\%$, $20\%$, $20\%$, $20\%$ and $20\%$, $10\%$, $20\%$, $10\%$, $20\%$, $20\%$ respectively while
the relative weights for the pre-test and post-test questions for class 3 were
$20\%$, $20\%$, $30\%$, $30\%$ and $30\%$, $20\%$, $30\%$, $20\%$ respectively.
For tutorial V, the relative weights for both the pre-test and post-test questions for class 2 were $30\%$, $40\%$ and $30\%$
and the weights for the post-test questions for the other two groups were $10\%$, $20\%$, $20\%$,
$20\%$, $10\%$, $20\%$ respectively. For all other cases, the same weight is assigned to each pre-test or post-test question.}
\label{junk2}
\end{table}


%% file: All_Rounded_new.tex



\begin{table}[h]
\centering
\begin{tabular}[t]{|c|c|c|c|c|c|c|c|c|c|c|c|}
\hline 
	N &	Tutorial&		Range ($\%$)&   n1 (class 1)&	pre&  post& n2 (class 2)&pre&post&n3 (class 3)&pre&post\\[0.5 ex]
\hline
	194&	 I&		All&     82&	53&	90& 60 &48 &96& 52&39&83 \\[0.5 ex]			
	   &      &     	0-33&    24&	19&	77& 21 &20 &92& 29&22&76 \\[0.5 ex]
	   &      &     	34-67&   33&	55&	92& 18 &43 &97& 21&58&92 \\[0.5 ex]
	   &      &     	68-100&  25&	83&	99& 21 &80 &99& 2 &100&100 \\[0.5 ex]					
\hline
	63 &	II&		All&     &	&	& 63 &35 &92& && \\[0.5 ex]			
	   &      &     	0-33&    &	&	& 30 &17 &89& && \\[0.5 ex]
	   &      &     	34-67&   &	&	& 32 &50 &94& && \\[0.5 ex]
	   &      &     	68-100&  &	&	& 1 &90 &90&  && \\[0.5 ex]					
\hline
	182&   III&		All&     78& 42	&81	& 55 &44 &85& 49&40& 85\\[0.5 ex]			
	   &      &     	0-33&    31&  18&76	&  22&20 &81& 22&15& 85\\[0.5 ex]
	   &      &     	34-67&   38& 52	&84	&  26&55 &87& 19&53& 85\\[0.5 ex]
	   &      &     	68-100&  9& 84	&88	&  7&79 &88&  8&78& 85\\[0.5 ex]					
\hline
	176&    IV&		All&     65&	40&	91& 62 &36 &89& 49&30&87 \\[0.5 ex]			
	   &      &     	0-33&    26&	15&	89& 31 &17 &85& 29&16&83 \\[0.5 ex]
	   &      &     	34-67&   32&	50&	91& 27 &51 &94& 17&47&92 \\[0.5 ex]
	   &      &     	68-100&  7&	83&	96& 4 &78 &94&  3&70&93 \\[0.5 ex]					
\hline
	57 &	 V&		All&     &	&	& 57 & 27&80& && \\[0.5 ex]			
	   &      &     	0-33&    &	&	& 42 & 14&77& && \\[0.5 ex]
	   &      &     	34-67&   &	&	& 10 &47 &90& && \\[0.5 ex]
	   &      &     	68-100&  &	&	& 5 & 96&92&  && \\[0.5 ex]					
\hline
\end{tabular}
\vspace{0.1in}
\caption{ Percentage average pre-/post-test scores (matched pairs) for each of the five tutorials (I-V), 
divided into three groups according to the pre-test performance. $N$ 
denotes the total number of
students who worked through a tutorial and took both the pre-/post-tests, and 
$n_i$ (i=1,2,3) denote the number of students in a particular class.
For tutorials II and V, only one of the classes took both the pre-/post-tests. For students in the high
pre-test range, sometimes there are very few students for a meaningful statistical interpretation.
}
\label{junk2}
\end{table}


%% file: Gauss_Law_Avg_short.tex
\begin{table}[t]
\centering
\begin{tabular}[t]{|c|c|c|c|c|c|c|}
\hline

&Without tutorial but otherwise & Honors Students&\multicolumn{2}{|c|}{Upper-level Undergrads}&With Tutorial&First Year Grads \\[0.5 ex]\cline{4-5}
&same type of courses (2 classes)& (2 classes) &Pre-test& Post-test& (4 classes)& (2 classes)\\[0.5 ex]
\hline
&	N=135 &	N=182 &	N=33&	N=28&	N=278 &	N=33 \\[0.5 ex]	
\hline
Average &	38 &	42&	44&	49&	59&	75 	\\[0.5 ex]
\hline
p value &	1.34E-04&	7.85E-04&	4.33E-03&	5.29E-02& &	1.37E-03 \\[0.5 ex]

\hline
\end{tabular}
\caption{ The average percentage of correct responses to the ``Superposition, Symmetry and 
Gauss's Law Test" for different groups of students. $N$ refers to the total number of students for a given group. In all undergraduate courses,
the test was administered after instruction on these concepts in that course except in the upper-level $E\&M$ course in which it 
was given both as a pre-test and post-test (since students had instruction in these concepts at the introductory level). 
The ``without tutorial" group and the ``Honors students"  group
are from the same student population (mainly physical science and engineering freshmen but some sophomores) as the tutorial group from the same institution.
The second row of the table gives the 
$p$ value for t-tests (which performed a pair-wise comparison
of the performance of the tutorial group with each of the other groups before rounding off the numbers).}
\label{junk2}
\end{table}